\documentclass{vgtc}                          

\ifpdf
  \pdfoutput=1\relax                   
  \usepackage{graphicx}                
  \DeclareGraphicsExtensions{.pdf,.png,.jpg,.jpeg} 
\else
  \usepackage{graphicx}                
  \DeclareGraphicsExtensions{.eps}     
\fi%

\graphicspath{{figures/}{pictures/}{images/}{./}} 

\usepackage{microtype}                 
\PassOptionsToPackage{warn}{textcomp}  
\usepackage{textcomp}                  
\usepackage{mathptmx}                  
\usepackage{times}                     
\usepackage{cite}                      
\usepackage{tabu}                      
\usepackage{booktabs}                  

\usepackage{url}

\onlineid{0}

\vgtccategory{Research}

\vgtcinsertpkg

\newcommand{\theName}{\textit{{WordStream Maker}}}

\title{WordStream Maker: A Lightweight End-to-end Visualization Platform for Qualitative Time-series Data}

\author{Huyen N. Nguyen\thanks{e-mail: huyen.nguyen@ttu.edu}\\ %
     \parbox{1.4in}{\scriptsize \centering University of New Hampshire \\ Texas Tech University}
\and Tommy Dang\thanks{e-mail: tommy.dang@ttu.edu}\\ %
     \scriptsize \centering Texas Tech University %
\and Kathleen A. Bowe\thanks{e-mail: kathleen.jeffery@unh.edu}\\ %
     \parbox{1.4in}{\scriptsize \centering University of New Hampshire}}

\teaser{
 \centering
 \includegraphics[width=\linewidth]{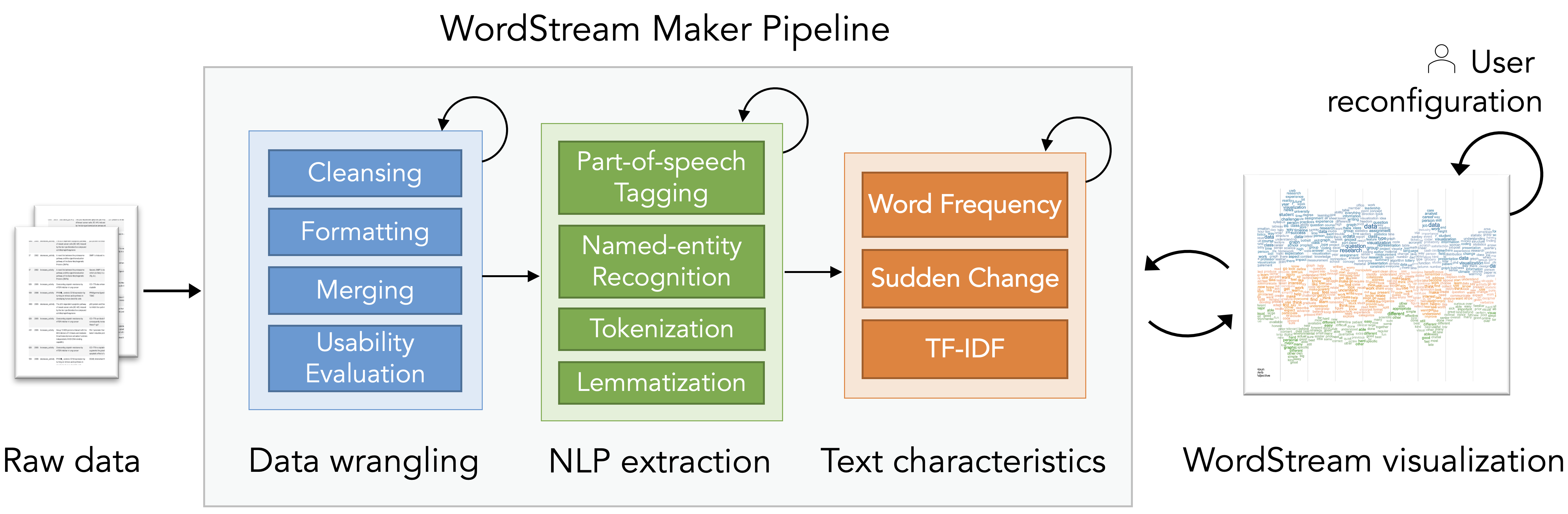}
 \caption{Architecture of the \textit{WordStream Maker,} with the integration of data wrangling, natural language processing (NLP) extraction and computing text characteristics for visualization.}
 \label{fig:pipeline}
}

\abstract{Whether it is in the form of transcribed conversations, blog posts, or tweets, qualitative data provides a reader with rich insight into both the overarching trends as well as the diversity of human ideas expressed through text. Handling and analyzing large amounts of qualitative data, however, is difficult, often requiring multiple time-intensive perusals in order to identify patterns. This difficulty is multiplied with each additional question or time point present in a data set. A primary challenge then is creating visualizations that support the interpretation of qualitative data by making it easier to identify and explore trends of interest.  By combining the affordances of both text and visualizations, WordStream has previously enabled ease of information retrieval and processing of time-series text data, but the data-wrangling necessary to produce a WordStream remains a significant barrier for non-technical users. In response, this paper presents \textit{WordStream Maker}: an end-to-end platform with a pipeline that utilizes natural language processing (NLP) to help non-technical users process raw text data and generate a customizable visualization without programming practice. Lessons learned from integrating NLP into visualization and scaling to large data sets are discussed, along with use cases to demonstrate the usefulness of the platform.  %
} 

\CCScatlist{
  \CCScatTwelve{Human-centered computing}{Visu\-al\-iza\-tion}{Visualization systems and tools}{};
  \CCScatTwelve{Human-centered computing}{Visu\-al\-iza\-tion}{Visu\-al\-iza\-tion application domains}{Information visualization};
}

\begin{document}

\maketitle

\section{Introduction}
From everyday communication to analytical conversations, qualitative text data is around us in all shapes and forms. Documents, news, and transcriptions of audio and video recordings are examples of qualitative data; an area that has gained huge growth in recent years–social media data–is also considered qualitative data on a quantitative scale~\cite{socialMedia}. With the rapidly increasing amount of data over time, the question is: how can we best leverage the data that we already have to deal with on a daily basis into a more valuable asset?

Our previous effort to convey the evolution of topics over time in multiple categories resulted in the WordStream visualization~\cite{dang2019wordstream}. WordStream combines the advantages of Wordle~\cite{viegas2009participatory} and stream graph~\cite{harris1999information} into a hybrid visualization to represent the topic streams, especially when they highly fluctuate. By preserving the terms in the visualization, WordStream enables both ease of information retrieval and information processing. However, the production of WordStream is still difficult for non-technical users. Furthermore, a recent work by Nguyen et al.~\cite{nguyen2021interactive}, indicates a gap between data visualization and qualitative research in data visualization literacy. Hence, our primary aim is to build a visualization platform that helps bridge this gap and ease the process of extracting insights from temporal patterns in text data.

Transforming typical raw data into usable data involves the process of cleansing, merging, formatting, extracting, and converting– generally known as 'data wrangling'~\cite{kandel2011research} that amounts up to 80\% of the development time and cost in data warehousing projects~\cite{dasu2003exploratory}. In qualitative data, this tedious task is coupled with the context-laden, conceptual characteristics of the data itself, making data processing increasingly challenging. With the advancement of natural language processing (NLP), the complex textual structure can be broken down and interpreted to assist humans in comprehension and communication. However, applying NLP to process text data remains challenging to non-technical users, especially when a large data set is involved. In \textit{WordStream Maker,} the processing of raw input data is embedded in the pipeline, and parameters are available for customization on demand.

\theName{} is inspired by several interesting existing works. First is the concept of the reusable chart, proposed by Mike Bostock~\cite{reusableChart}, which provides reconfiguration of a visualization model for customization flexibility depending on the user's needs. Following this concept, RAWGraphs by Mauri et al.~\cite{mauri2017rawgraphs} introduces a visualization platform to create open outputs, opening many possibilities for encoding data dimensions on uncommon visual models. For static word clouds, the Word Cloud Generator developed by Jason Davies~\cite{wordcloudGenerator} offers a straightforward way to generate such clouds efficiently. However, there is still a lack of platforms for text data and qualitative data in general, especially text data coupled with the time element. 

Therefore, we created an end-to-end platform to help process raw time-series text data on the fly, extracting text characteristics with NLP, generating the time-series visualization, and aiding users in customizing the representation. The system is lightweight and runs completely on the end user's browser; no server side and database needed also help with data privacy. The lessons learned from integrating NLP directly into visualization and scaling to large data sets are also discussed.

The rest of the paper is structured as follows: Section~\ref{relatedWork} presents related work to \theName{}. Section~\ref{platform} describes \theName{} in more detail in terms of architecture and user interface. Section~\ref{usecases} discusses the findings with \theName{}. Finally, Section~\ref{conclusion} concludes this paper and gives an outlook for future work.

\begin{figure}[t]
 \centering 
 \includegraphics[width=\columnwidth]{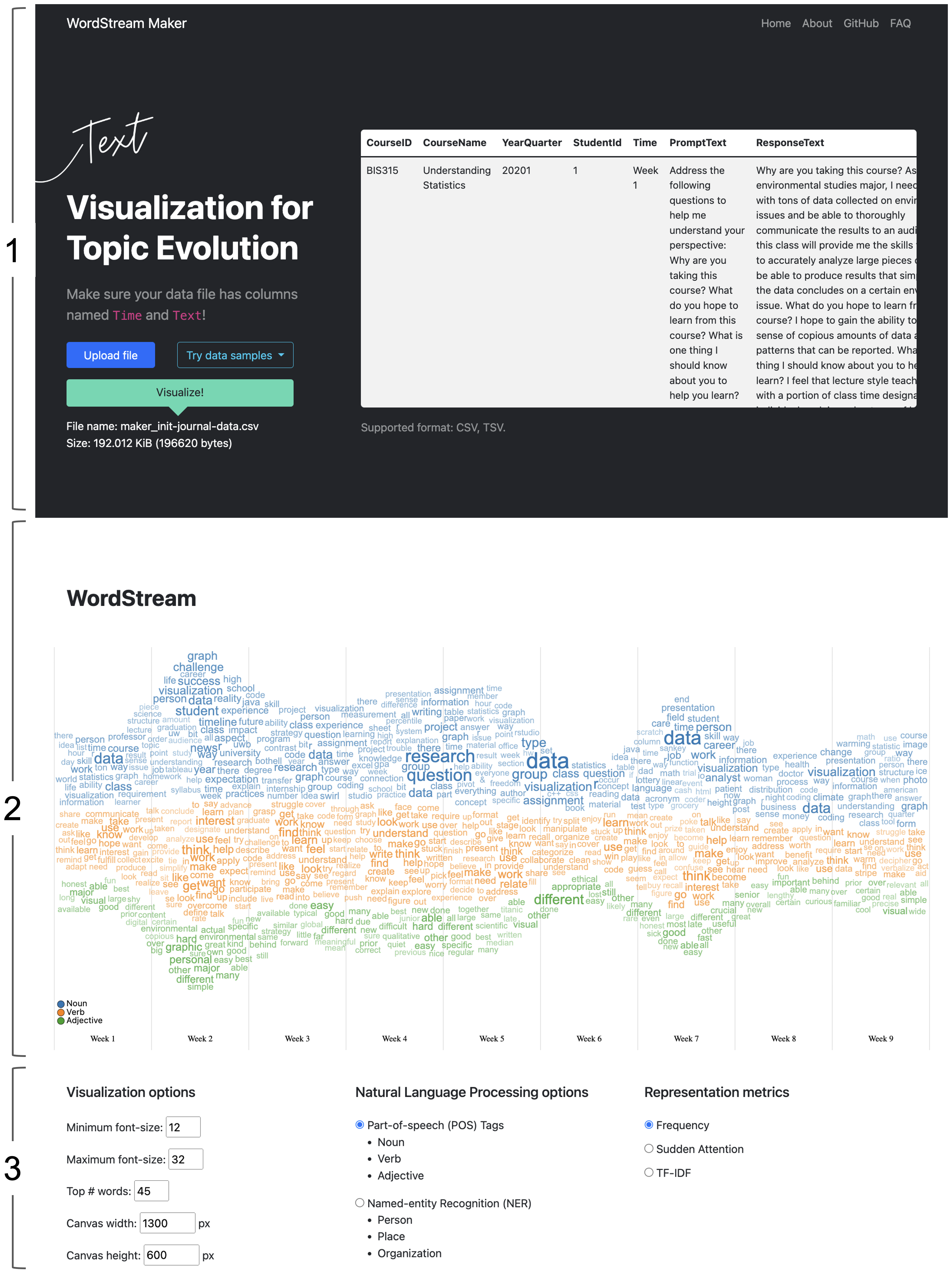}
 \caption{Visual interface of \theName{}.}
 \label{fig:interface}
\end{figure}

\section{Previous Work}
\label{relatedWork}
In the state-of-the-art report on text data streams by Wanner et al.~\cite{wanner2014state}, word clouds are considered to be the only visualization technique based on textual data, while timeline and circular representations are more prevalent to visualize the evolution of the data over time. A more recent work by Ottley et al.~\cite{ottley2019curious} indicates that effective representations should take advantage of the affordances of both text and visualization. Taking this into account, we developed WordStream~\cite{dang2019wordstream} as an effort to represent the topic streams in a temporal manner.

The technique of WordStream has been applied in various domains to explore the evolution of text data over time. The visualization was first introduced with topics from IEEE VIS publications, Huffington Post collection, along with other online blogs~\cite{dang2019wordstream}. During the COVID-19 pandemic, the visualization is integrated into a tool called CovidStream to monitor the evolution of emotions associated with COVID-19 in Peru by Baca et al.~\cite{baca2020covidstream}. The technique is also employed to analyze emerging topics on social media: assisting earthquake situational analysis~\cite{nguyen2019eqsa, chen2019earthquakeaware}, cybersecurity~\cite{van2019hackernets} and natural resources~\cite{dang2020agasedviz}. Recent work shows the extensions of WordStream on educational assessments~\cite{nguyen2021interactive} and in the search for potential crowdfunding investors~\cite{zhang2022will}.

Several directions are proposed to incorporate NLP into visual analytics. A natural language interface (NLI) in interactive visualization systems can offer graphical answers to address ambiguous questions from users~\cite{hoque2017applying, setlur2016eviza}. Narechania et al.~\cite{narechania2020nl4dv} introduced NL4DV, a robust Python package assisting visualization developers in creating new visualization NLIs and integrating natural language input into their systems without prior knowledge in NLP. While training and task framing remain challenges due to the characteristics of natural language input~\cite{srinivasan2020ask}, by applying automated text mining methods in extracting and discovering knowledge in unstructured data sources~\cite{basole2013innovation}, there are many open opportunities in the intersection of natural language and visualization.

\section{\theName{} Platform}
\label{platform}
\theName{} is an open source web application\footnote{\url{https://github.com/huyen-nguyen/maker}} based on mainly two libraries: D3.js~\cite{bostock2011d3} for the creation of WordStream visualization and Compromise~\cite{compromise} library for NLP engine.

\subsection{Platform architecture}
The overall architecture of \theName{} platform is presented in Figure~\ref{fig:pipeline}. First, the primary input to the platform is a raw time-series text file. Input data then goes through the main component of \textit{WordStream Maker,} consisting of three sub-components: data wrangling, NLP extraction, and text characteristics computing. As the data source is heterogeneous, data cleansing is crucial to ensure the data is credible and usable. There are many cases where more than one row shares the same timestamp; hence merging data is the next important step. 

The next sub-component is the NLP engine. This module provides Part-of-speech (POS) tagging, e.g., labeling whether a word is a noun, a verb, or other parts of speech such as preposition or adverb. POS tagging not only helps to extract the desired category but also makes it easier to remove stop words, which are commonly used and often bear little meaning. These stop words are often determiners, conjunctions, or prepositions; hence, we can skip these categories by using POS tagging. Diving in the Named-entity Recognition (NER) within the Noun category, three popular categories are chosen: Places, Person, and Organization. The application of these NLP modules in visualization will be discussed in Section~\ref{usecases}. NLP engine also provides lemmatization, helping to trace the root form of any given word. This function is very useful in WordStream visualization; for example, the verb ``study'' in the present tense (study/studying) or past tense (studied) conveys the same topic; hence should be represented in the same root word ``study''.

The last sub-component is text characteristics computing, which serves directly to the visualization. The importance of a term can be represented by frequency, as in Wordle and other word cloud formations. In \textit{WordStream Maker,} we use two other characteristics, namely sudden change and term frequency-inverse document frequency (TF-IDF). While TF-IDF is a statistical measure that evaluates the relevance of a word to a document in a collection of documents~\cite{qaiser2018text}, sudden change is calculated based on sharp changes in frequency. Let $F_1, F_2, \ldots, F_n$ be the frequency of a word at $n$ different time points. The sudden change series ($S_1, S_2, \ldots, S_n$): $S_t= \frac{(F_t+1)}{(F_{t-1}+1)}$.

\subsection{User interface}

The user interface of \theName{} is divided into three sections, as shown in Figure~\ref{fig:interface}: (1) Data loading and preview, (2) WordStream visualization, and (3) Options for customization.

In the first section, users have the option of uploading a file containing tabular data in the tabular format of comma-separated values (CSV) or tab-separated values (TSV). For qualitative time-series data, the file must have one column representing the time element and one representing the body of text. There is also an option to load a sample data set for demonstration purposes.  After the data is loaded, the window on the right will render a preview of input data as a table.

The second section is the WordStream visualization. At this point, the text data has been completely processed and parsed, and keywords have been extracted. The WordStream library will generate the visualization based on extracted data, using a default configuration. The width of the stream presents the total frequency, and the font size of each word corresponds to its individual frequency (or other selected text characteristics).

The last section is for customizing the view: any adjustments will directly affect the WordStream visualization above. This section is divided into three groups: visualization-related, NLP-related, and text representation-related. In the first group, users can adjust the minimum and maximum font size in the WordStream, the number of words displayed in each stream for each time step, and the sizes (width and height) of the view. In the second group, the user can select to use POS tagging (default) or NER. POS tagging offers three categories: Noun, Verb, and Adjective, while NER corresponds to three main sub-categories: Person, Place, and Organization. In the last group, representation of text characteristics, the three options are frequency (default), sudden change, and TF-IDF.

\subsection{Availability}
\theName{} is available at \url{https://huyen-nguyen.github.io/maker/}.

\section{Discussion}
\label{usecases}

\subsection{Use cases: Educational Assessment Data}

The educational assessment data set~\cite{nguyen2021interactive} contains 63 entries with attributes such as \textit{Course ID}, \textit{Course Name}, \textit{Prompt Text}, \textit{Response Text}. This data set reflects the journal entries where students give answers to weekly prompts made by instructors for class assessments. In this case, \theName{} is used for exploration without specific visual analytics tasks.

\begin{figure}[t]
 \centering 
 \includegraphics[width=\columnwidth]{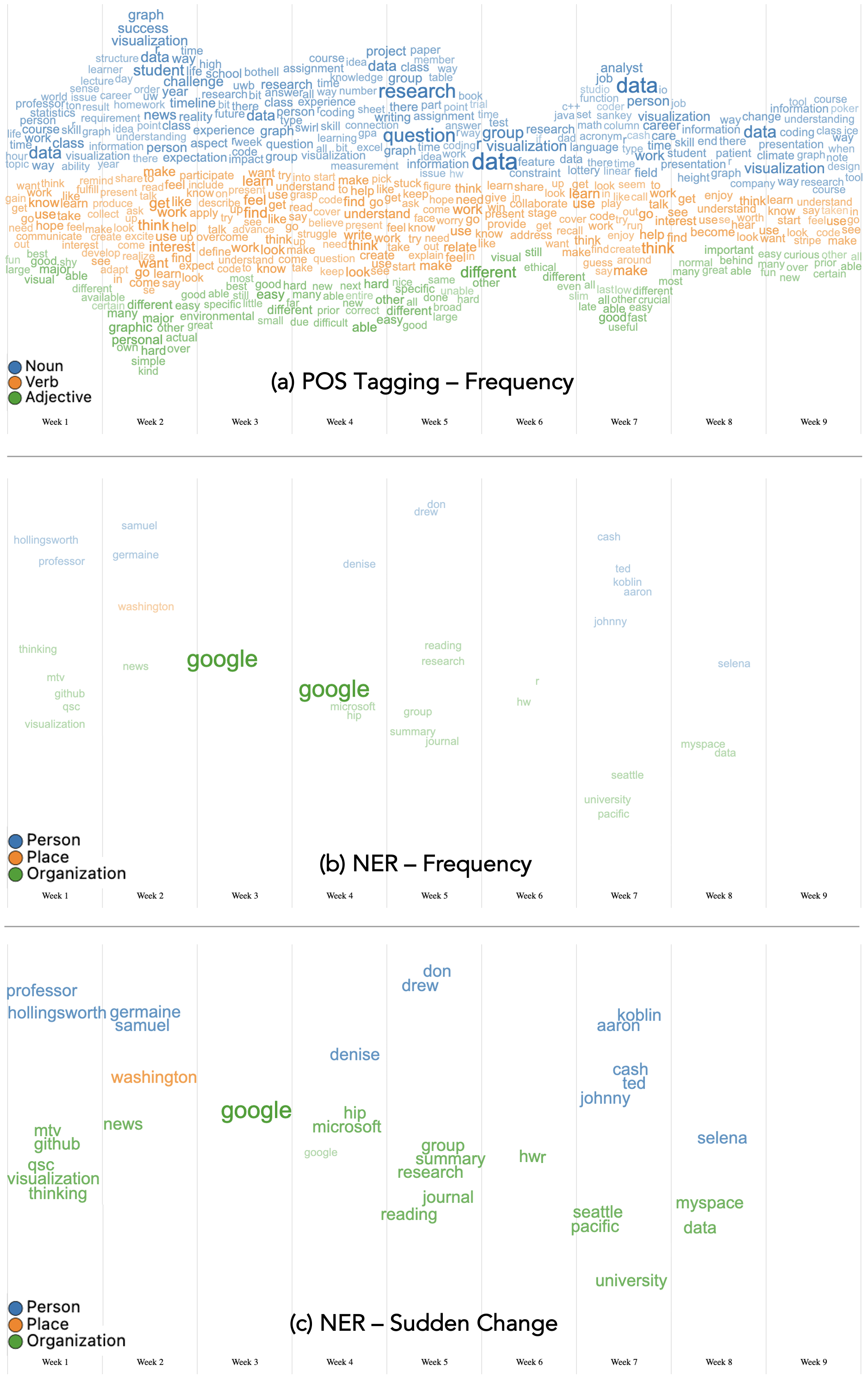}
 \caption{Journal data for education assessments throughout nine weeks. Different combinations of Part-of-speech (POS) Tagging, Named-entity Recognition (NER) in the NLP module, along with Frequency and Sudden Change in the text characteristics module.}
 \label{fig:student}
\end{figure}

With POS tagging and frequency option (Figure~\ref{fig:student}(a)), words are classified by Noun, Verb, and Adjective, with their font size indicating their occurrences in the text source. When we keep the frequency option and change POS tagging to NER (Figure~\ref{fig:student}(b)), we can observe that: 1) the number of words falling into the new categorization greatly declines, and 2) there are only two prominent terms in the second view and they are of the same word ``google''. This is because POS tagging provides more generic categorization, and the only organization that gains the most interest (based on frequency) is Google. When we keep the NER categorization and change to sudden change representation (Figure~\ref{fig:student}(c)), now the minor words in the previous view have the chance to shine, as there is a sharp increase in the mention of other organizations such as ``microsoft'', ``github'', ``myspace''. This type of observation is useful in data exploration, especially in the early stage of the analysis process when the users may not be sure about what questions should be asked.

\subsection{Data exploration}
\begin{figure}[t]
 \centering 
 \includegraphics[width=0.9\columnwidth]{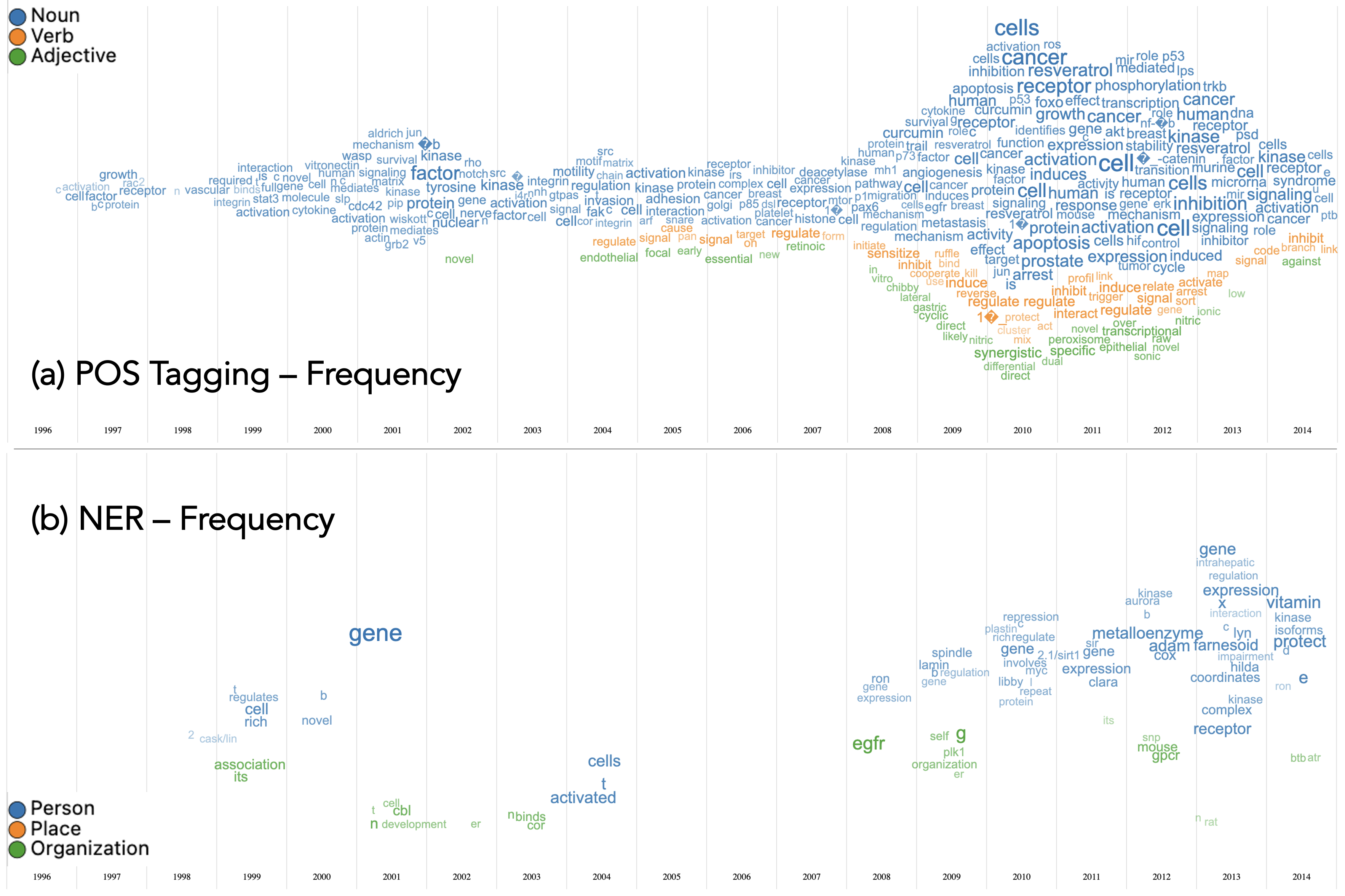}
 \caption{Pathway protein. Different combinations of Part-of-speech (POS) Tagging and Named-entity Recognition (NER) with frequency.}
 \label{fig:p}
\end{figure}

For different domains and different tasks, different ways of categorization should be applied to the qualitative data. Another use case is presented in Figure~\ref{fig:p}: Keywords from publications on protein pathways from 1996 to 2014. Summarizing from previous study~\cite{nguyen2019eqsa, dang2020agasedviz, nguyen2021interactive} and \textit{WordStream Maker,} we observe that: 1) NER categorization works best for data from social media or written form of response. 2) POS tagging can be employed for various domains; however, its generalization can be too broad to form a coherent story or specific insight.

\subsection{Trade-offs}

With the purpose of building a lightweight application, and the data processing part is meant to be on the fly, the selection on the NLP engine itself requires a modest-sized library. We go with Compromise due to the small size of its minified version (250kb) and the diverse NLP functionalities it provides.
When it comes to tokenization– break a chunk of words into smaller units, there are two options. One is to break the sentence into individual words: a word is determined whenever we encounter the next whitespace. The other way is to break sentences into noun chunks, each containing one or more words; this way, we can preserve the meaning of the noun phrases. While the former approach is more straightforward, the second approach helps to maintain the semantics of text structure. 

We run the experiment on a macOS Monterey 12.2.1 of 2.5GHz, and 16GB RAM. In runtime, for a text file of size 1.5MB, containing more than 5000 rows, the first approach takes less than 3 seconds to load data, clean data, and run NLP extraction. However, the second approach takes 1 minute and 50 seconds, which is very time-consuming. With the aim of a lightweight platform, we choose simple tokenization at a small cost of semantics.

\section{Conclusion and Future Work}
\label{conclusion}
In this paper, we present \textit{WordStream Maker,} a lightweight web-based platform to help non-technical users process raw text data and generate a customizable visualization of WordStream without programming practice. The application is open source and meant to be expandable. Outlook for future work will focus on supporting more input types, user interactions, and applicability of \theName{} in a variety of use cases.

\acknowledgments{
This project was supported by the National Aeronautics and Space Administration (NASA) and the Gordon Research Conferences under NASA Award \#17-TWSC17-0055. This paper was completed during the first author's internship at the University of New Hampshire, whose warm hospitality and support were greatly appreciated.}

\bibliographystyle{abbrv-doi}

\bibliography{references}
\end{document}